# Social Networks and Construction of Culture: A Socio-Semantic Analysis of Art Groups[1]

Nikita Basov, Ju-Sung Lee, and Artem Antoniuk

## 1 Introduction

Network analysts combining culture and networks have shown that culture is linked to social relations. One the one hand, researchers argue that culture is reproduced through interactions and therefore relies on concrete interpersonal ties (e.g., [7, 8, 36]). On the other hand, it is shown that culture affects structure of social ties (e.g., [23, 14]. In sum, culture and social networks are seen as mutually constitutive, or dual [5, 25].

Most of the above studies view culture as a set of constructs which combine ideas, concepts, and meanings shared among individuals (for an overview, see [25]). These constructs correspond to similar ways of interpreting the world and condition similarities in preferences, tastes, ideas, and judgments (for an overview, see [30]). Cultural constructs are exhibited in verbal (written or spoken) expressions of people who belong to the same culture [26, 38]. In these expressions, structures of associations between words rather than the words themselves represent cultural meanings [33, 37]. Hence, research has been advocating a structural view on verbally expressed culture [6, 9]. Yet, the relations between social networks and culture as *structure* have not been sufficiently analyzed, especially in small groups (see in relevant overviews by [25] and [5]). This paper investigates ***how social network positions of actors in the social networks associate with cultural constructs they create jointly with other group members.***

Using semantic network analysis based on word collocation [34, 6, 32, 16], we trace cultural constructs as patterns of associations between concepts expressed by individuals' and relate the properties of those cultural constructs to positions in net-

Nikita Basov and Artem Antoniuk
Centre for German and European Studies, St. Petersburg State University – Bielefeld University, St. Petersburg, Russia, e-mail: n.basov@spbu.ru and e-mail: artiom.a000@gmail.com

Ju-Sung Lee
Erasmus University Rotterdam, Rotterdam, The Netherlands, e-mail: lee@eshcc.eur.nl

works of social ties occupied by individuals. Hence, we apply the growingly popular socio-semantic framework [32, 31, 29, 27].

In particular, we focus on groups of visual artists. These groups jointly generate culture, most often in observable processes of creating corporeal artistic objects and group interactions, exchanging on – often joint – artwork creation, collective exhibitions, discussions on the events and figures of the artistic scenes, and other artistic and everyday topics. Network analysis has been widely applied to study creativity and social relations between artists. Yet, network studies of art have focused primarily on organizational and market levels (e.g., [13, 15, 35, 3]), while creativity is seen as dependent on an individual's [21, 11] position in a network of *external* relations [20]. The question of how internal networks of art groups operate appears to be out of scope. Meanwhile, it is those internal networks of art groups that bring to life novel artistic visions and artistic styles many groups strive for [17] thus generating variations in culture. So, it makes sense to take a closer look at such internal social networks. Simultaneously, research on language use has been argued to be "a powerful way to study the collective action of cultural production in art worlds" [12, p. 201]. Developments in semantic networks allow for the exploration of relations between cultural production and social networks within art groups. Yet, so far, very few studies applied formal semantic network analysis techniques to artistic settings [1]. This paper deals with this gap.

## 2 Data

The empirical data used in this study covers 3 art groups from St. Petersburg, Russia, encoded as 'A', 'B', and 'C'. All of them are working in the format of contemporary visual art. They all are characterized by intense interaction between the members, (decades-long) backgrounds shared by most of the members, and regular joint artistic and/or everyday practices. Hence, their cultural constructs may both affect their interactions and be impacted by these. Besides, the groups actively produce texts and narratives that can be used to capture expressed cultural constructs. Simultaneously, the groups are different in organization, educational and cultural backgrounds of their members, understandings of art and its tasks, forms of spatial embeddedness in the city space, and artistic styles. This provides variability in cases.

We collected data between 2011 and 2012 via in-depth ethnographic studies conducted in each of the 3 groups. Because the groups do not have formal boundaries, we decided to include only core members in the data collection, that is those members with stable membership and continuous involvement in the group practice.

The data consists of two main parts: textual data and sociometric data. The textual data includes verbal expressions of the group members with clearly identifiable individual authorship. The corpus of texts is composed of transcripts of 24 open-ended narrative interviews, each 30–240 minutes long, transcripts of dialogues between group members coming from 17 ethnographic observations, each 2–8 hours long, as well as posts in Russian social media, textual works of the artists, such as newspaper articles, prose and poetry. We managed to gain texts by every core member in all the 3 groups. Unprocessed individual corpora sized between 4128 and 28928 words per member.

The sociometric data was obtained using the roster recall method surveys capturing frequency of interactions among members of each group. The question asked was "How often do you interact?", suggesting to choose from 5 response options to evaluate frequency of interactions with each of other members of the group: almost never; 1 or less/month; 2–4 times/month; 5–14 times/month; 15 or more times/month. Further, responses were quantified on ordinal levels, from 0 for 'almost never' to 4 for '15 or more times/month'. 25 out of 29 core members responded to the survey resulting in a response rate of 86.21%.

## 3 Method

### *3.1 Mapping of the Socio-Semantic Networks*

To capture patterns of social ties, cultural structures, and structure of relations between them, three types of networks were mapped using the data on the three art groups: actor-actor (social network representing structure of social ties), concept-concept (semantic network representing cultural constructs) and actor-concept (bimodal concept usage network representing links between individuals and certain cultural constructs). Combined, these three types of networks constitute socio-semantic networks [32].

The edge widths of the actor-actor (social) networks in Fig. 1 are based on ordinal levels from 0 to 4 captured by the sociometric survey. Tie strength was taken as an average of individuals' evaluations of frequency of interactions with each other. When no response was received from one of the individuals in a dyad, only the strength indicated by the other one served as an input for the social network.

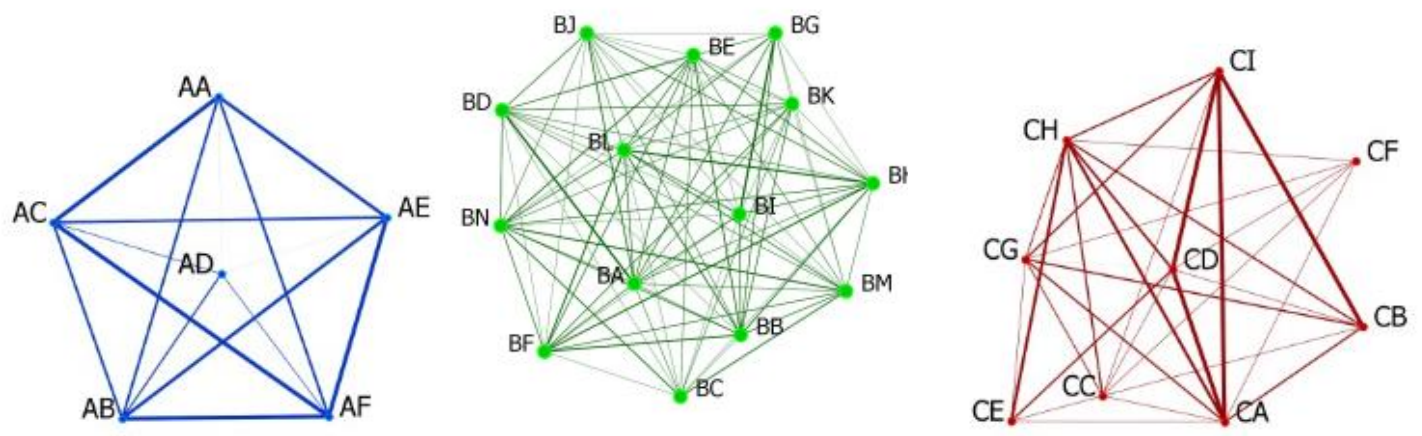

(a) **A** (b) **B** (c) **C**

**Fig. 1** Social networks of the 3 art groups: "**A**", "**B**", and "**C**".

The other two types of networks were mapped based on the collected texts. Concepts, which are stems of words used in texts, are the nodes in semantic and actor-concept networks. To map relations between concepts in semantic networks we used words collocation technique, which implies that links between nodes are mapped based on word stems co-occurrences in texts. These networks represent cultural constructs expressed by individuals [6, 25]. Relations between concepts and actors were mapped based on usage of certain concepts by certain individuals in their texts. Neither frequency of words' collocation nor frequency of words use were accounted for in this analysis, so both semantic and concept usage networks are binary.

The procedure for mapping semantic and concept usage networks was as follows. First, the textual data were split into separate files containing all narratives and written texts by each single group member, separately. Then, we removed interviewers' and observers' comments and technical information. Second, textual data were preprocessed in AutoMap [10], applying concept stemming, lowercasing, removal of punctuation and numerals. A delete list was created and applied, removing pronouns, adverbs, prepositions, conjunctions, junk words, as well as less meaningful verbs, such as 'say', 'talk', and 'think'.

Third, AutoMap was applied to each of these separate files to generate *individual semantic networks* of each artist. Parameters of semantic network generation were specified as follows: window size between 2 and 3 words was used to map lines between concepts; sentence was used as a stop unit.

Fourth, individual networks of each group member were aggregated into *union semantic networks* (so that links are now based on collocation of concepts in texts of any of the artists), while actor-concept networks still contained the information on usage of certain concepts by certain actors.

Fifth, concepts used by only one group member (i.e. having fewer than 2 binary actor-concept links) were removed from semantic and actor-concept networks as we are interested only in capturing shared cultural constructs. Therefore, our analysis includes only concepts used by at least two actors in a group. We note, however that in this paper, links between concepts are not necessarily shared.

The three types of networks (social, semantic, and bipartite concept usage) were mapped for each of the three art groups, resulting in 12 networks in total and comprising 3 socio-semantic networks of the 3 groups further used in this analysis.

### 3.2 Operationalization of the Social Network

The social network survey recorded the frequency of interactions among members of each group. However, this frequency was measured on ordinal levels, which carry concerns over numerical comparisons from one level to the next. For example, the ratios among levels differ from any estimated levels. So, we instead replace tie strengths with estimations of the actual frequency of contact.

Table 1 enumerates the estimates and ranges (for sampling) for tie strength ordinal scale values. In our subsequent analyses, the estimates, rather than the survey responses, are employed. An alternative approach is to consider uniform sampling of tie strengths using the estimated min and max actual frequencies (also shown in the tables). Finally, asymmetric interaction reports are symmetrized by averaging the dyadic reports.

**Table 1** Mapping of Tie Strengths to Ranges and Estimates

| Survey Response | Description | Min. | Max. | Estimate |
|---|---|---|---|---|
| 0 | Almost Never | 0.01 | 0.1 | 0.05 |
| 1 | 1 or less/month | 0.1 | 1.0 | 0.5 |
| 2 | 2–4 times/month | 1.5 | 4.5 | 3.0 |
| 3 | 5–14 times/month | 4.5 | 14.5 | 9.5 |
| 4 | 15 or more times/month | 14.5 | 20.0 | 20.0 |

### 3.3 Descriptive Statistics

We present some descriptive statistics of the social and semantic networks in Table 2. These groups are relatively small and may be considered to be small social networks, which bear the characteristic of being socially cohesive. That is, the social networks are highly dense. By contrast, the semantic networks exhibit extremely low densities, which is largely due to the high number of concept nodes and the co-word window employed in the semantic network generation. Despite the huge difference between the densities of social and semantic networks, we note that these densities are ordered similarly, suggesting a relation between social networks and cultural constructs. For example, the "**C**" network exhibits both the lowest concept and actor network densities.

**Table 2** Social and semantic network statistics

| | A | B | C |
|---|---|---|---|
| **Actors** | 6 | 14 | 9 |
| **Ties** | 15 | 89 | 28 |
| **Ord. Weighted Ties** | 44.5 | 152 | 53 |
| **Est. Wgt. Ties** | 163.75 | 284 | 141.75 |
| **Interactions/Tie** | 10.92 | 3.19 | 5.06 |
| **Social Network Density** | 1.000 | 0.978 | 0.778 |
| **Concepts** | 7513 | 4800 | 13681 |
| **Semantic Network Density** | 0.00077 | 0.00058 | 0.00039 |

### 3.4 Extraction of socio-semantic subgraphs

In Fig. 2, we visualize the union graph of the bipartite concept usage network and the unimodal social network for the "**C**" group. The concept usage network is optimized using the pivot multidimensional scaling (MDS) algorithm [4], as implemented in Pajek [2], so that structural equivalence is optimally displayed. That is, nodes that are connected to similar others are placed in proximity to one another and nodes connected to the same other nodes – exactly upon each other, thereby reducing the visual complexity induced by the 13,681 observed concepts. Clusters of concepts form distinguishable groups or 'bands' of concept nodes scattered around nodes of actors using them (See Fig. 3). The added value of such an optimization is that it gives a picture of how actors are grouped together with regard to usage of similar sets of concepts and concepts are grouped with regard to their usage by certain sets of actors.

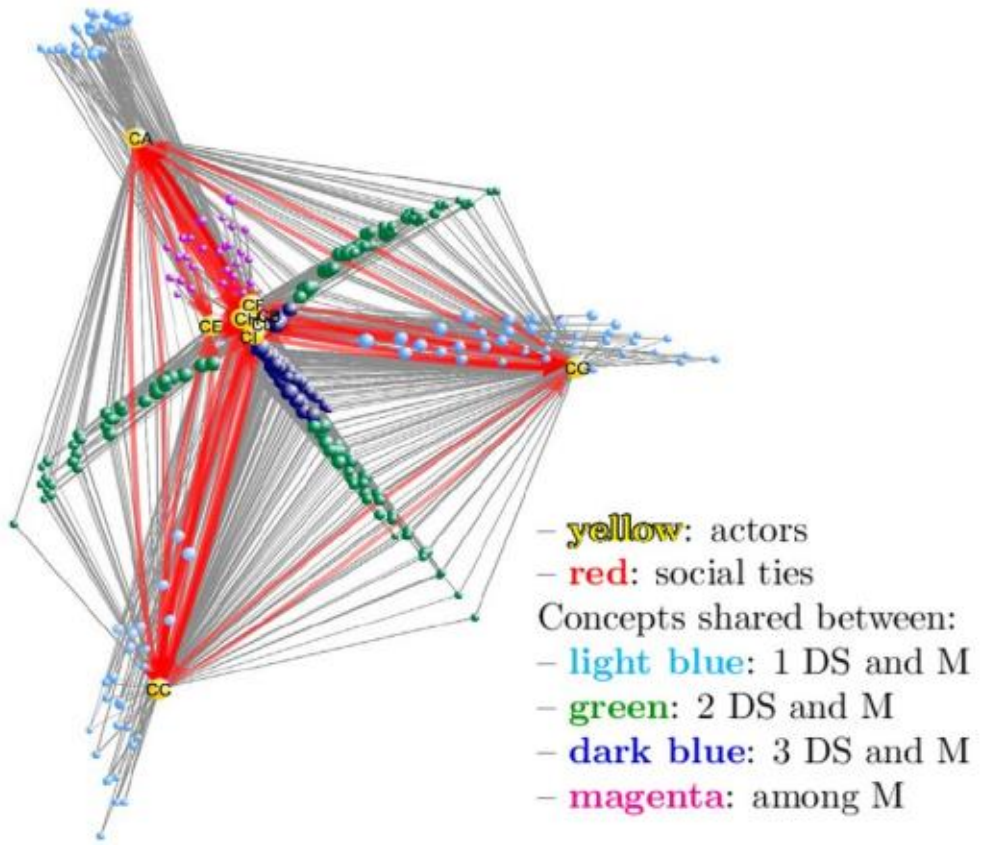

**Fig. 2** Visualization of actor-concept and actor-actor networks of group '**C**'.

Actor nodes are in yellow and labeled (anonymously). Concept nodes are colored according to combinations of actors sharing them and sized by the number of structurally-equivalent concepts. Grey lines refer to concept usages, while overlaid red lines represent social ties.

Due to the nature of pivot MDS algorithm, some actor nodes usually form a triangle-shaped structure with other actor nodes located in the middle of the diagram. Actor nodes located at triangle's vertices represent people who use the largest amount of concepts shared with other members, while actor nodes located inside represent those who use a significantly smaller number of shared concepts. Thus, a distinction between different positions of actors in the concept usage network is captured. Actor nodes located a triangle's vertices appear to be very different in their cultural constructs with regard to each other, as reflected by the content of the concepts they use. What they have in common is that they use many concepts, which are also used by many others in the group. Hence, they span the semantic space playing an important role in culture constructing in the group. Therefore, we label them 'discourse spanner(s)' (or **DS**). Simultaneously, these individuals appear to be informal leaders in their groups, acknowledged as such by other members and demonstrating corresponding behavior in group interactions. Due to their strong involvement in the formation of their groups' shared semantics, characteristics of the semantic networks that DS contribute to are most worth considering in order to understand how culture is constructed in groups.

The second type of position in the socio-semantic network is represented by the 'majority' (or **M**) of other actors who are using shared concepts to a much lesser extent and hence are less involved in culture constructing within their groups. The positions of DS and M represent not only two different types of positions in the concept usage structure, but also two distinctive roles in group culture constructing corresponding to these positions. Although in this paper we mainly focus on the DS, we still account for M.

As an important technical step, for each group we extracted different socio-semantic subgraphs. These subgraphs include (1) different combinations of DS and M actors, (2) concepts shared by them, as suggested by the pivot MDS optimization (e.g., blue concepts in Fig. 2 correspond to concepts shared by 1 of the DS and one or more of the M), and (3) any links between concepts they have in their semantic networks; we note that links between concepts are not necessarily shared.

Fig. 3 represents an example semantic network of a socio-semantic subgraph which includes concepts used by the DSs, encoded as CC and CG, share with some of the M (green nodes in Fig. 2). It represents, for instance, that the concepts 'poem' and 'prose' are used by DS 'CC', DS 'CG' and at least one individual in M; meanwhile, the association represented by the link 'poem'-'prose' may be characteristic of only CC, or CG, or the individual(s) from among the M. Due to the limits of space, our analysis in this paper considers only those socio-semantic subgraphs that include one DS and one or more of the M.

## 4 Results

As a starting point, in Table 3, we predict (in the statistical, non-causal sense) concept usage by individual actors' social network position statistics, namely degree ($C_D$) and betweenness ($C_B$) centralities.[3] The former measure captures the extent to which an actor interacts with others, while the latter indicates the extent to which an actor plays a bridging role within his/her group. [18, 19]. We examine centrality measures derived from the undirected, unweighted graphs as well as the estimated, empirical edge weights, in order to address homogeneity of unweighted degree centralities due to high density in some groups. The models are applied across all groups (total of 29 members), and the dependent variable is log-transformed due its skewness.

The results primarily reveal that betweenness ($C_B$) is positively associated to shared concept usage by individuals while degree (or popularity, $C_D$) is negatively associated. The relative, absolute magnitudes of these effects vary by the opera-

---

[2] Node size corresponds to betweenness centrality of concepts. Pendants were recursively hidden in the main picture. The full semantic network is displayed in the lower-right.

[3] Due to the low sample size, we cannot include additional predictors or employ a nested model. However, group size, while significant on its own, is collinear with $C_D$, but does not predict as well not do group-level dummy variables.

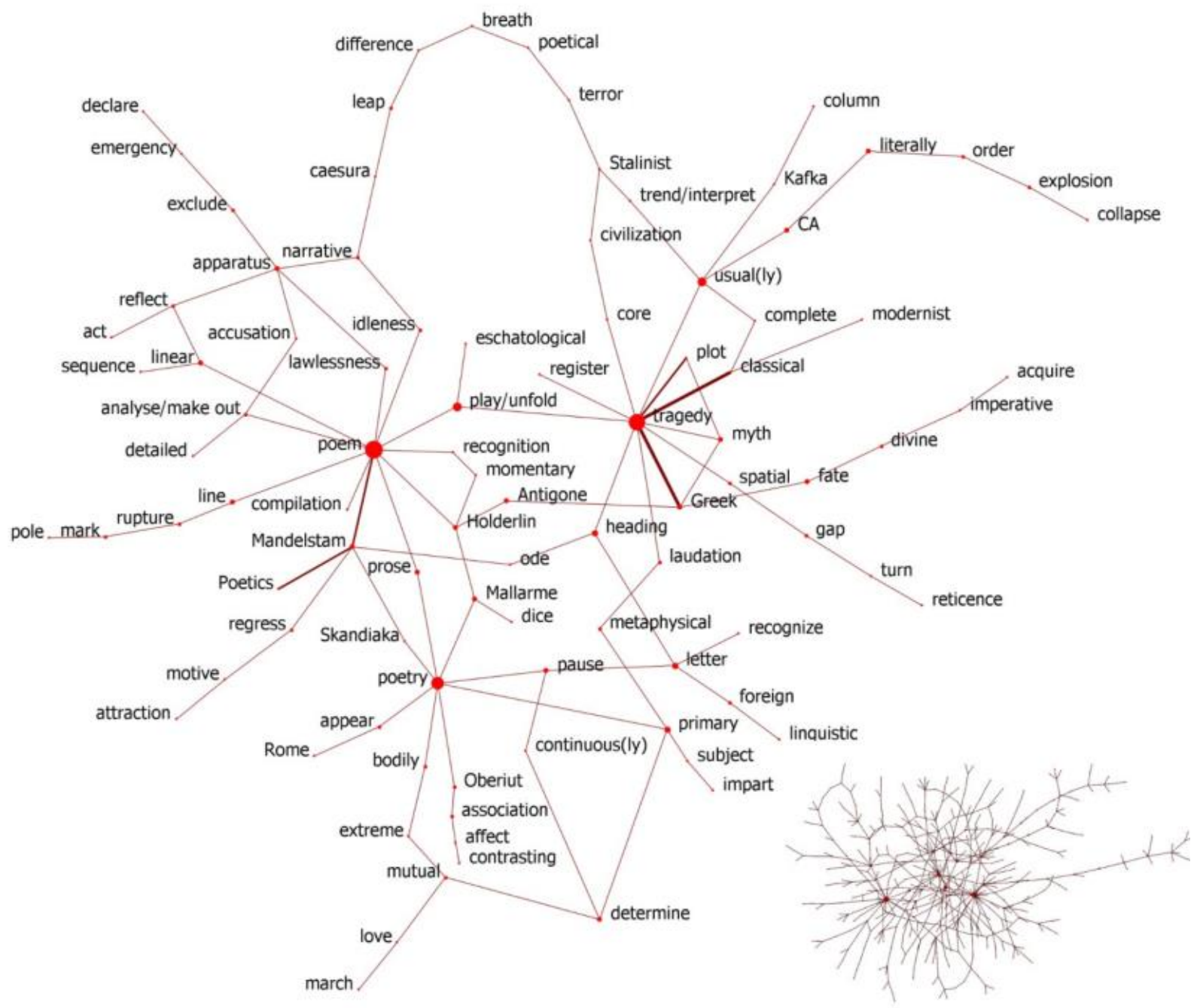

**Fig. 3** Largest pruned component of the semantic network of a socio-semantic subgraph with CC, CG, and M.[2]

**Table 3** Predicting concept usage by social network measures

| Predictor | log of Shared Concept Usage | *Wgted. Centrality Predictors* |
|---|---|---|
| Intercept | 7.992*** | 7.728*** |
| | (0.429) | (0.500) |
| $C_D$ | −0.165*** | −2.156* |
| | (0.043) | (0.825) |
| $C_B$ | 0.642* | 1.258* |
| | (0.241) | (0.604) |
| Adj-$R^2$ | 0.423 | 0.162 |
| $n$ | 29 | 29 |

$*: p < .05; ***: p < .001$

Note: Second model uses the same dependent variable as the first but alternative, weighted predictors.

tionalization of the tie, whether it is mere existence or the strength.[4] That is, actors use and share with others more concepts when they connect areas of their social networks, while they use and share fewer concepts when they intensely interact with their closer circles. This reveals that conceptual prominence of DS is hindered by their popularity but is empowered by their ability to connect the group. Given that degree and betweenness are often positively correlated, the negative effect of $C_{\mathrm{D}}$ reveals that those DS who use particularly many concepts are rather distinctive gatekeepers than merely merely globally central through high ranking on both measures [28, 29, 27].

In Table 4, we compare the weighted social network to the 'concept sharing network'; the latter is the bipartite concept usage network transformed via network multiplication (or folding) into a unimodal actor-actor network in which the edge weight represents the extent of shared concepts. For each group and socio-semantic subgraph corresponding to different types of roles (DS and M), the correlations between the edges of the social network and those of the concept sharing network are tested for significance under a permutation test that produces a null distribution resistant to type I errors induced by matrix (network or distance) auto-correlation [24]; the resulting correlation is called a 'QAP correlation' [22]. We examine the relationship between the social ties and concept sharing by pairs of actors for each group, in general, as well as for DS and M subgraphs within each group. These subgraphs strictly contain only social ties among DS (or M, respectively) and the concepts DS (or M) share between them. The edge weights in the concept sharing networks are additionally log-transformed due to skewness.

**Table 4** QAP correlations between shared concepts and social ties per subgroup

| Name | Pearson $r$ | Pearson $r$ (w/log trans.) | $n$ |
|---|---|---|---|
| *Discourse Spanners (DS) Subgraph* | | | |
| **A** | — | — | 2 |
| **B** | $.298^{n.s.}$ | $.296^{n.s.}$ | 4 |
| **C** | $.345^{n.s.}$ | $.399^{n.s.}$ | 3 |
| *Majority (M) Subgraph* | | | |
| **A** | $.041^{n.s.}$ | $.035^{n.s.}$ | 4 |
| **B** | $.123^{n.s.}$ | $.128^{n.s.}$ | 10 |
| **C** | $-.337^{\wedge}$ | $-.401^{\wedge}$ | 6 |
| *All Members Graph* | | | |
| **A** | $-.034^{n.s.}$ | $-.001^{n.s.}$ | 6 |
| **B** | $.034^{n.s.}$ | $.105^{n.s.}$ | 14 |
| **C** | $-.284^{\wedge}$ | $-.350^{*}$ | 9 |

*n.s.* : $p \geq .10$; $^{\wedge}$ : $p < .10$; $*$ : $p < .05$

[4] Results from considering weighted concept usage (multiple use per concept by a single individual) are very consistent with the shown results.

Despite the small samples, there are some results worth mentioning. First, we see that the DS social and concept sharing networks (for those groups that contain more than two DS) exhibit positive, albeit insignificant, correlations. These suggest the social ties between DS as a subgroup and concept sharing between them have an ambivalent association: either strong ties act as a normalizing force on inducing a common dictionary or vice versa.

The M subgraphs also exhibit this ambivalence with the exception of group **C**, whose significantly negative correlation indicates that the more strongly M actors are tied, the fewer concepts they share. This suggests that certain M members of group **C** sought distinction from one another in their cultural constructs. This correlation remains when we look at the entire **C** group despite the normalizing nature of the DS of that group.

As the above analysis shows, mere concept sharing by individuals does not demonstrate any prominent relation with social ties. Further, we account for cultural meanings by considering links between concepts (semantic networks), and we search for relations between social ties linking actors and cultural constructs the actors jointly express. We compare semantic network statistics of subgraphs that include one DS and the M (i.e. separate union semantic networks connecting concepts shared by DSs and one or more of Ms) against normalized sum of weights (i.e. the sum of dyadic degree centralities divided by maximum sum possible) of the edges a DS has with the M. Specifically, we compare graph-level measures (GLMs) computed for the semantic networks (per sets of DS+M) against the interaction strength the DS exhibits with the M in their respective groups. This comparison (Table 5) exposes the extent to which the cultural constructs created by discourse spanners together with the majority of other actors in their groups relate to cumulative strength of social ties between the DS and the M within a group.

**Table 5** Semantic network statistics v. averaged social network tie strengths

| Measure | $r_{\text{ord}}$ | $r_{\text{est}}$ | $\bar{r}_{\text{MC}}$ |
|---|---|---|---|
| Density | $.116^{n.s.}$ | $-.047^{n.s.}$ | $-.001^{n.s.}$ |
| Degree Centralization | $.375^{n.s.}$ | $.259^{n.s.}$ | $.314^{n.s.}$ |
| Betweenness Centralization | .883** | .822** | .814* |

$n.s.: p \geq .10, *: p < .05, **: p < .01$

There are $n = 9$ DS (3, 2, and 4 for each group respectively). Density indicates the unweighted density of the semantic networks in each DS+M socio-semantic subgraph. Degree and betweenness centralizations are variance-based metrics of the distribution of nodal degree and betweenness centralities and are normalized between 0 and 1. They reveal the extent to which the structure contains concepts that a) harbor significantly more semantic linkages to other concepts and b) play prominent bridging roles in the semantic network, connecting disparate areas of a group's cultural constructs and thus integrating them. Together, they can describe properties of cultural constructs. For example, higher density would suggest 'thickness' of cultural constructs; and lower degree centralization may indicate more diversified cultural constructs, in contrast to those that are narrowly focused.

We report the nominal Pearson correlations (*r*) derived from both the ordinal responses and empirical estimates as well as a mean from Monte Carlo sampled tie strengths; these are all consistent with one another for the higher correlations. Significant and positive correlations for betweenness centralization indicate that the presence of distinct concepts that prominently bridge semantic networks accompanies stronger bonds between a DS and a M. In other words, strong social ties are associated with integration of cultural constructs. The other positive correlation, for degree centralization, although insignificant, points towards decreased diversification of cultural constructs as being associated with stronger social ties between a DS and a M. It suggests that the cumulatively strong ties between DS and M may make group discourse elaborate on some narrow set of focal concepts, perhaps mobilizing the group discourse. Thus, focusing and, especially, integration of cultural constructs rather than mere 'thickness' of cultural constructs accompany intense interactions between DS and the M.

## 5 Conclusion

This study focused on relating social networks and cultural constructs in art groups, with implications on social and cultural duality extending to other domains. By studying the interplay between social and semantic networks, we attempted to shed light on the relation social role and position of an individual have with his/her involvement in constructing shared culture in a group. Minding that our findings are limited due to analysis of cross-sectional data on small groups embedded in a single – artistic – setting, we can summarize the following.

First, the analysis revealed that, even in small groups of friends, higher diversity and intensity of direct social ties hinders sharing of cultural constructs. Rather, those individuals, who socially bridge less well-connected areas of their groups, are the ones who engage in the shared cultural constructs with others.

Second, the amounts of concepts shared by the group members and strength of social ties between them are not necessarily related. While those individuals using significant amounts of shared concepts bear some of this association (which Roth and Cointet refer to as "semantic homophily" [32], in one of the groups, the members employing significantly fewer shared concepts exhibit heterophily, whereby stronger ties are marked by lower levels of concept sharing. This finding differs from those of [32] (ibid.) which, however, rely on analysis of much larger groups.

Finally, we found that stronger focusing and higher integration of cultural constructs rather than mere 'thickness' of cultural constructs accompany more intense interactions between the leaders and the followers. Our preliminary interpretation is that leaders are strategically interacting with others in order to jointly construct a shared creative vision and to integrate the group. In this process, leaders rely not only on their competence or formal authority, but also on focusing on emerging cultural constructs and on interaction with others. The more intensely they interact with the rest of the group, the more they bridge and focus the individual group's cultural constructs on a shared set of concepts serving to span the group's culture. At the moment, we cannot say for sure whether or not it is a phenomenon specific to creative settings, and if there is an asymmetric relation. This issue will be addressed in our analysis of longitudinal data, currently being collected.

Overall, we can preliminarily conclude that the socio-semantic network approach is capable of delivering findings on the duality of cultural and social structures relevant to the ongoing discussion (see [25, 5] Yet, the analysis would benefit from a more extensive account of links between concepts (semantic networks) and from combining of quantitative and qualitative data. We expect that joint formal analysis of semantic network properties with contents of semantic networks, along with ethnographic and interview data, will deliver further insights.

**Authors' Contributions** The first two co-authors contributed equally to this paper and are listed alphabetically.

**Acknowledgements** The paper has benefited from the support of: St. Petersburg State University ("Communication practices of knowledge creation in the social space of a contemporary city", 2011-2012), Russian scientific foundation for humanities (15-03-00722 "Coevolution of knowledge and communication networks: structural dynamics of creative collectives in European cultural capitals", 2015–ongoing), and the Centre for German and European Studies Bielefeld University and St. Petersburg State University supported by the DAAD with funds from the German Foreign Office. Also, the authors express their gratitude to those who helped in data collection and processing: Aleksandra Nenko, Anisya Khokhlova, Elena Tykanova, Maria Veits, Olga Volkova, Irina Shirobokova, Alexey Evstifeev, Alexander Kopiy, and Olga Nikiforova. We would also like to thank Wouter de Nooy, who proposed to use pivot MDS optimization for the given data, and Adina Nerghes for her comments and logistical assistance in assembling this paper. We are also very grateful for comments received on this paper from Iina Hellsen.

## References

1. Basov, N., Nenko, A.: Artistic community knowledge structure revealed: A semantic network analysis of 'La Escocesa', Barcelona. Centre for German and European Studies pp. 3–32 (2013)
2. Batagelj, V., Mrvar, A.: Pajek-program for large network analysis. Connections **21**(2), 47–57 (1998)
3. Bottero, W., Crossley, N.: Worlds, fields and networks: Becker, Bourdieu and the structures of social relations. Cultural Sociology **5**(1), 99–119 (2011)
4. Brandes, U., Pich, C.: An experimental study on distance-based graph drawing. In: 16th International Symposium on Graph Drawing, vol. 5417, pp. 218–229. Springer-Verlag (2008)
5. Breiger, R.L., Puetz, K.: Culture and networks. In: International encyclopedia of social and behavioral sciences (2nd. ed ed.). Elsevier, New York (2015)
6. Carley, K.: Extracting culture through textual analysis. Poetics **22**(4), 291–312 (1994)
7. Carley, K.M.: An approach for relating social structure to cognitive structure. Journal of Mathematical Sociology **12**(2), 137–189 (1986)
8. Carley, K.M.: A theory of group stability. American Sociological Review **56**(3), 331–354 (1991)

9. Carley, K.M.: Extracting team mental models through textual analysis. Journal of Organizational Behavior **18**(1), 533–558 (1997)
10. Carley, K.M., Columbus, D., Landwehr, P.: Automap User's Guide 2013. Tech. rep., CMU-ISR-CASOS, Pittsburgh, PA (2013)
11. Cattani, G., Ferriani, S.: A core/periphery perspective on individual creative performance: Social networks and cinematic achievements in the Hollywood film industry. Organization Science **19**(6), 824–844 (2008)
12. Cluley, R.: Art words and art worlds: The methodological importance of language use in Howard S. Becker's "Sociology of art and cultural production". Cultural Sociology **6**(2), 201–216 (2012)
13. Crane, D.: The transformation of the avant-garde: The New York art world, 1940-1985. University of Chicago Press, Chicago, IL (1989)
14. Dahlander, L., McFarland, D.A.: Ties that last: Tie formation and persistence in research collaborations over time. Administrative Science Quarterly **58**(1), 69–110 (2013)
15. De Nooy, W.: A literary playground: Literary criticism and balance theory. Poetics **26**(5), 385–404 (1999)
16. Diesner, J.: From texts to networks: Detecting and managing the impact of methodological choices for extracting network data from text data. KI-Knstliche Intelligenz **27**(1), 75–78 (2013)
17. Farrell, M.P.: Collaborative circles: Friendship dynamics and creative work. University of Chicago Press, Chicago, IL (2003)
18. Freeman, L.C.: A set of measures of centrality based on betweenness. Sociometry **40**(1), 35–41 (1977)
19. Freeman, L.C.: Centrality in social networks: Conceptual clarification. Social Networks **1**(3), 215–239 (1979)
20. Guimera, R., Uzzi, B., Spiro, J., Amaral, L.A.N.: Team assembly mechanisms determine collaboration network structure and team performance. Science **308**(5722), 697–702 (2005)
21. Hargadon, A., Sutton, R.I.: Technology brokering and innovation in a product development firm. Administrative Science Quarterly **42**, 716–749 (1997)
22. Krackhardt, D.: QAP partialling as a test of spuriousness. Social Networks **9**, 171–186 (1987)
23. Lizardo, O.: How cultural tastes shape personal networks. American Sociological Review **71**(5), 778–807 (2006)
24. Mantel, N.: The detection of disease clustering and a generalized regression approach. Cancer Research **27**(2), 209–220 (1967)
25. Mohr, J.W.: Measuring meaning structures. Annual Review of Sociology **24**, 345–370 (1998)
26. Mohr, J.W., Duquenne, V.: The duality of culture and practice: Poverty relief in New York City, 1888–1917. Theory and Society **26**(2), 305–356 (1997)
27. Nerghes, A., Hellsten, I., Groenewegen, P.: A toxic crisis: Metaphorizing the financial crisis. International Journal of Communication **9**, 106–132 (2015)
28. Nerghes, A., Lee, J.S., Groenewegen, P., Hellsten, I.: The shifting discourse of the European Central Bank: Exploring structural space in semantic networks. In: K. Yetongnon, A. Dipanda (eds.) Proceedings of the 10th SITIS, pp. 447–455. IEEE Computer Society (2014)
29. Nerghes, A., Lee, J.S., Groenewegen, P., Hellsten, I.: Mapping discursive dynamics of the financial crisis: A structural perspective of concept roles in semantic networks. Computational Social Networks **2**(16) (2015)
30. Pachucki, M.A., Breiger, R.L.: Cultural holes: Beyond relationality in social networks and culture. Annual Review of Sociology **36**(1), 205–224 (2010)
31. Roth, C.: Socio-semantic frameworks. Advances in Complex Systems **16**, 0405 (2013)
32. Roth, C., Cointet, J.P.: Social and semantic coevolution in knowledge networks. Social Networks **32**(1), 16 – 29 (2010)
33. Saussure, F.D., Hidayat, R.S.: Pengantar Linguistik Umum. Gadjah Mada University Press, Yogyakarta, Indonesia (1988)
34. Sinclair, J.: Corpus, concordance, collocation. Oxford University Press, Oxford (1991)
35. Uzzi, B., Spiro, J.: Collaboration and creativity: The small world problem. American journal of sociology **111**(2), 447–504 (2005)

36. White, H.C.: Identity and control: A structural theory of social action. Princeton University Press, Princeton, NJ (1992)
37. Wittgenstein, L.: Philosophical Investigations (Vol. 50). Prentice Hall, New York (1953)
38. Yeung, K.T.: What does love mean? Exploring network culture in two network settings. Social Forces **84**(1), 391–420 (2005)